\documentclass[reprint,
 amsmath,amssymb,
 aps,
prstab
]{revtex4-1}

\usepackage{graphicx}% Include Figure files
\usepackage{subcaption}
\usepackage{stackengine}
\usepackage{dcolumn}% Align table columns on decimal point
\usepackage{bm}% bold math
\usepackage{xcolor}
\usepackage{comment}
\usepackage{amsmath}

\usepackage{textcomp} %Textmu

\makeatletter
\newcommand\thefontsize[1]{{#1 The current font size is: \f@size pt\par}}
\newcommand\thefontsizeHere{{The current font size is: \f@size pt\par}}
\makeatother

\begin{document}

\preprint{}

\title{Beam Dynamics Framework Incorporating Acceleration to Define the Minimum Aperture in Two Focusing Schemes for Proton Radiotherapy Linac}% Force line breaks with \\
%\title{An investigation of rf instabilities during beam loading of recirculating Energy Recovery Linac}% Force line breaks with \\

\author{M.~Southerby}
\email{m.southerby@lancaster.ac.uk}
\author{R.~Apsimon}
\email{r.apsimon@lancaster.ac.uk}
\affiliation{Engineering Department, Lancaster University, Lancaster, LA1 4YW, UK }
\affiliation{Cockcroft Institute, Daresbury Laboratory, Warrington, WA4 4AD, UK}

\date{\today}% It is always \today, today,

\begin{abstract}

In this paper, a self-consistent transverse beam dynamics framework is demonstrated, that incorporates acceleration into the transverse beam dynamics studies for a proton linac machine. Two focusing schemes are developed and discussed; the FODO-like scheme, and the minimum aperture scheme. The FODO-like scheme is a simple scheme, requiring only one quadrupole per cavity. The scheme is analytically solved to minimise the beam size at the cavity entrance/exit and ensures a constant beam size along the lattice, with respect to adiabatic damping due to longitudinally accelerating rf cavities. The minimum aperture scheme describes the regime that matches the beam ellipse to the acceptance ellipse of a cavity, allowing for the smallest possible aperture, for a given cavity length. A simple approximation of an rf cavity map is determined to allow changes in particle energy along a lattice, and acceleration is assumed only in the longitudinal direction.

\end{abstract}

\maketitle

\section{Introduction}

In the recent decades, all-linac solutions for proton acceleration with medical applications have become an increasing area of interest \cite{future_linacs}. Two areas to have benefited from such improvements are cancer radiotherapy and medical imaging \cite{radio, imaging}. All-linac solutions have benefits over the conventional cyc-linac and synchrotron machines with respect to energy and intensity modulated on the scale of ms. This allows for more efficient treatment of cancers with proton beams, such as active spot scanning for moving organs \cite{pbs}. In addition to advantages to radiotherapy, linac boosters can be used in conjunction with cyc-linac or all-linac solutions to push proton energy to 350 MeV, the energy required for medical imaging \cite{imaging_proton}. Proton medical imaging allows a more accurate calculation of the required proton energy during radiotherapy, over conventional X-ray imaging, due to the proton stopping power.

All-linac machines benefit from a smaller beam emittance than cyc-linac machines, and therefore can operate with smaller beam apertures, increasing the shunt impedance. Limits are often placed on the beam aperture due to the transverse focusing requirements of the linac, in addition to peak fields and power coupling.

This paper describes the method used to minimise the beam aperture with respect to transverse beam losses, for a given cavity length, analytically. The paper will discuss two focusing schemes, namely the FODO-like scheme, and the Minimum Aperture Scheme (MAS), incorporating longitudinal momentum gain. The FODO-like scheme is similar to the well-known FODO scheme, comprised of quadrupole of alternating polarity to produce a net focusing force transversely. The MAS scheme produces a matching section that aligns the transverse beam ellipse with the cavity acceptance ellipse. An RF cavity transfer map is produced to simulate longitudinal acceleration of protons, and the corresponding adiabatic damping that occurs as a result. Due to the very low beam currents used in proton radiotherapy linacs (of the order nA), space-charge effects are ignored. 

The Twiss parameter transfer matrix is adapted to account for the change in beam emittance due to acceleration. The method requires minimising the Twiss beta function, $\beta$, at the cavity entrance and exit to minimise the beam aperture for a given cavity length and beam emittance, whilst ensuring maximum beam acceptance.

\section{\label{sec:level1}Twiss Parameters with Acceleration}

The phase space ellipse of a particle in a periodic beam line, with geometric emittance $\varepsilon_{g, x}$, is described;
\begin{equation}
    \varepsilon_{g, x} = \beta_x x'^2 + 2\alpha_x x x' + \gamma_x x^2.
    \label{eq:beam_ellipse}
\end{equation}
Where $\beta_x$, $\alpha_x$ and $\gamma_x = \frac{1+\alpha_x^2}{\beta_x}$ are the Twiss parameters in $x$ \cite{wolski}. $x$ is the transverse size of the beam, and $x' = \frac{dx}{ds}$ = $\frac{p_{x}}{p_z}$ for longitudinal displacement, $s$. The maximum beam size at any point $s$, is given $\sigma = x_{max} =  \sqrt{\beta(s) \varepsilon_g(s)}$.

It is required to use the Lorentz invariant normalised emittance, defined as $\varepsilon_{n} = \varepsilon_g(s) \gamma_r(s)\beta_r(s)$ \cite{wolski}. Where $\gamma_r(s), \beta_r(s)$ are the Lorentz factor and normalised particle velocity, respectively. Using Eqn.~\ref{eq:beam_ellipse} to equate the normalised emittance of a particle before and after an rf cavity;
\begin{equation}
  \begin{aligned}
    \gamma_{r0}\beta_{r0}(\beta_{x1} x_1'^2 + 2\alpha_{x1} x_1 x_1' + \gamma_{x1} x_1^2) = \\
    \gamma_{r1}\beta_{r1}(\beta_{x0} x_0'^2 + 2\alpha_{x0} x_0 x_0' + \gamma_{x0} x_0^2)
  \end{aligned},
  \label{eq:equal_norm_emit}
\end{equation}
where $\gamma_{r0}\beta_{r0}$ and $\gamma_{r1}\beta_{r1}$ are the Lorentz factor and normalised particle velocity at the start and end of the cavity, respectively. The cavity can be described with a linear transfer map, $R$.
\begin{equation}
    \begin{pmatrix}
    x_{1}\\
    x'_{1} 
    \end{pmatrix}
    = 
    \begin{pmatrix}
    R_{11} & R_{12} \\
    R_{21} & R_{21}
    \end{pmatrix}
    \begin{pmatrix}
    x_{0} \\
    x'_{0}
    \end{pmatrix}.
    \label{eq:simple_map}
\end{equation}
Assuming only longitudinal acceleration, the divergence before and after the cavity is given;
\begin{equation}
    x_1' = \frac{\Delta p_x + x_0' p_{z0}}{p_{z1}},
    \label{eq:x_1'}
\end{equation}
where $\Delta p_x$ can be determined from Lorentz force, for a particle of charge $q$, longitudinal velocity $\beta_z c$, and at azimuthal angle, $\theta$;
\begin{equation}
    \Delta p_x = q \cos\left(\theta\right)\left(\frac{\int E_r dz}{\beta_{rz} c} + \int B_{\theta} dz\right).
    \label{eq:delta_px}
\end{equation}
In an azimuthally symmetric cylindrical cell, the radial electric field ($E_r$) and azimuthal magnetic field ($B_\theta$) can be written as functions of the longitudinal electric field, $E_z$, using a first order expansion about $r = 0$;
\begin{equation}
     E_r  = -\frac{r}{2} \frac{dE_z}{dz}, \hspace{0.5cm}
    B_{\theta} = \frac{\omega r}{2c^2}  E_z.
    \label{eq:ErBtheta}
\end{equation}
Where $\omega$ is the angular frequency.
A typical $E_z$ field component can be written as a Fourier series \cite{matt_fc2ct}, with the most simple case being;
\begin{equation}
    E_z = \sin\left(\frac{\pi z}{L_{cell}}\right)\sin\left(\omega t+ \phi_0\right)
    \label{eq:simple_Ez},
\end{equation}
for a given cell length, $L_{cell}$. $\phi_0$ represents the mean phase over the cavity. The value of $E_z$ as observed by a particle at constant velocity can be determined by substituting $t= \frac{z}{\beta_z c}$ into Eqn.~\ref{eq:simple_Ez}.

Using equations~\ref{eq:simple_Ez}, \ref{eq:ErBtheta}, \ref{eq:delta_px} and \ref{eq:x_1'} produces an approximation for $x_1'$;
\begin{equation}
    x_1' = \frac{N \pi}{4p_{z1}}\left(\beta_{rz0} - \frac{1}{\beta_{rz0}}\right) \sin(\phi_0)x_0 + \frac{p_{z0}}{p_{z1}}x_0'.
    \label{eq:defoc_mapping_x'}
\end{equation}
Where $N$ is the number of rf cells in the rf cavity, and $N L_{cell} = L_{cav}$. Integrating Eqn.~\ref{eq:defoc_mapping_x'} over the cavity length produces a similar form for $x_1$. The final result of the cavity map is shown below;

\begin{equation}
    \begin{pmatrix}
    x_{1}\\
    x'_{1} 
    \end{pmatrix}
    = 
    \begin{pmatrix}
    1 + \frac{N \pi}{4}\left(\beta_{rz0} - \frac{1}{\beta_{rz0}}\right) \sin(\phi_0) L' &  \gamma_{r0}\beta_{rz0}mc L' \\
     \frac{N \pi}{4\gamma_{rs}\beta_{rz1}mc}\left(\beta_{rz0} - \frac{1}{\beta_{rz0}}\right) \sin(\phi_0) & A_d
    \end{pmatrix}
    \begin{pmatrix}
    x_{0} \\
    x'_{0}
    \end{pmatrix}.
    \label{eq:cavity_mapping_full}
\end{equation}
Where,
\begin{multline}
     L' = \frac{L_{cav}}{\Delta \gamma \cos(\phi_0) mc }\biggl( \cosh^{-1}({\gamma_{r0} + \Delta \gamma \cos(\phi_0)}) - \\ \cosh^{-1}(\gamma_{r0})\biggr),
     \label{eq:Lprime_defoc}
\end{multline}
with $\Delta \gamma = \gamma_{r1} - \gamma_{r0}$, and
\begin{equation}
    A_d = \left({{1 + \Delta \gamma_r \cos(\phi_0) \frac{\Delta \gamma_r \cos(\phi_0) + 2 \gamma_{r0}}{\gamma_{r0}^2 -1} }}\right)^{-1/2}.
\end{equation}
To proceed, the rf phase is chosen such that longitudinal acceleration is maximised, and defocusing forces are minimised,  $\phi_0 = 0$, as these are the conditions of the ideal particle
\begin{equation}
    \begin{pmatrix}
    x_{1}\\
    x'_{1} 
    \end{pmatrix}
    = 
    \begin{pmatrix}
    1&   L_{cav}\frac{\gamma_{r0}\beta_{rz0}}{\gamma_{r1} - \gamma_{r0}}ln\left(\frac{\gamma_{r1}\beta_{rz1} + \gamma_{r1}}{\gamma_{r0}\beta_{rz0} + \gamma_{r0}}\right)  \\
    0 & \frac{\gamma_{r0}\beta_{rz0}}{\gamma_{r1}\beta_{rz1}}
    \end{pmatrix}
    \begin{pmatrix}
    x_{0} \\
    x'_{0}
    \end{pmatrix}.
    \label{eq:cavity_map_simple}
\end{equation}

\begin{comment}
\begin{figure*}
    \centering
    \includegraphics[width=12cm]{figures/pz+dpz.png}
    \caption{Momentum vector of a particle before (red) and after (blue) a longitudinal momentum boost $dp_z$.}
    \label{fig:my_label}
\end{figure*}
\end{comment}
The Twiss parameter transfer matrix can be derived by substituting $x_0, x_0'$ as functions of $x_1, x_1'$ (using the inverse form for Eqn.~\ref{eq:simple_map}) into Eqn.~\ref{eq:equal_norm_emit}.

\begin{multline}
    \begin{pmatrix}
    \beta_{x1}\\
    \alpha_{x1} \\
    \gamma_{x1}
    \end{pmatrix} = \\ \frac{\gamma_{r1}\beta_{r1}}{\gamma_{r0}\beta_{r0}}
    \begin{pmatrix}
    R_{11}^2 & -2R_{11}R_{12} & R_{12}^2 \\
    -R_{11}R_{21} & R_{11}R_{22} + R_{12}R_{21} & -R_{12}R_{22} \\
    R_{21}^2 & -2R_{21}R_{22} & R_{22}^2 
    \end{pmatrix} \\
    \begin{pmatrix}
    \beta_{x0}\\
    \alpha_{x0} \\
    \gamma_{x0}
    \end{pmatrix}.
    \label{eq:twiss_matrix}
\end{multline}
The Twiss parameter transfer matrix takes on the recognised form for zero acceleration, when $\frac{\gamma_{r1}\beta_{r1}}{\gamma_{r0}\beta_{r0}} = 1$, as expected. 
\begin{comment}
As all mapping matrices in a FODO except for a cavity have unit determinant, the determinant of the half FODO, det(M) is simply the determinant of the cavity map, exploiting;
\begin{equation}
    det(ABC) = det(A)det(B)det(C)
\end{equation}
\end{comment}

The basic start point has now been established, with a simple cavity transfer map, and the Twists parameter transfer matrix incorporating acceleration.

\section{\label{sec:level1}FODO-like Scheme}

The aim of the FODO-like scheme is to analytically provide the quadrupole $k$-strength and length such that the beam size is minimised at the cavity entrance/exit, producing the smallest beam aperture possible for a given chosen cavity length.

\begin{figure*}
    \centering
    \includegraphics[width=16cm]{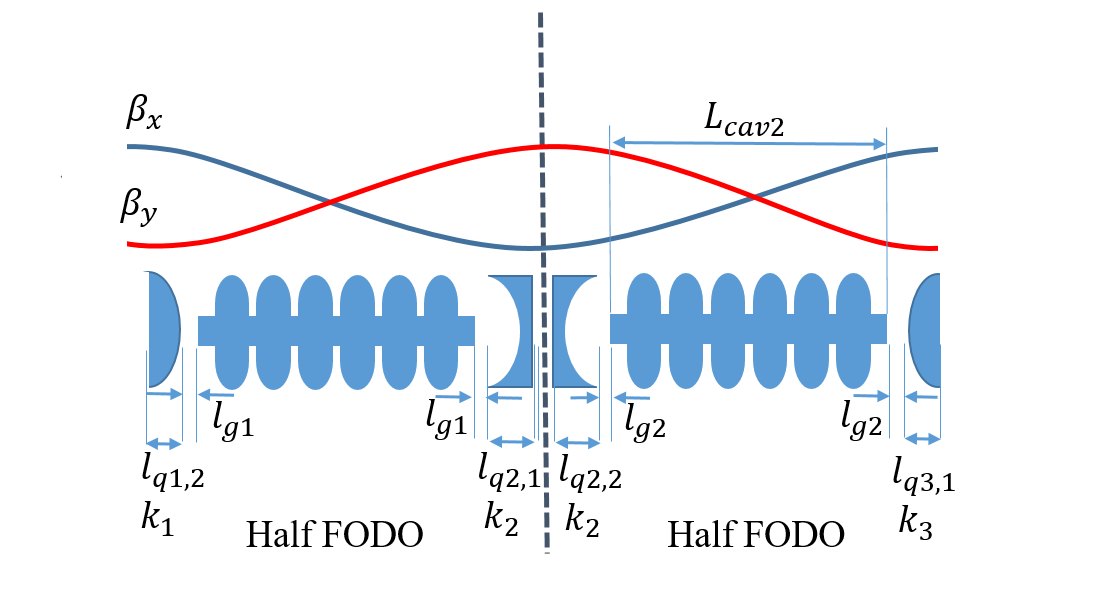}
    \caption{FODO-like schematic.}
    \label{fig:fodo_schem}
\end{figure*}

The lattice starts at a location such that the Twiss alpha function is $0$ in both transverse planes;
\begin{equation}
    \alpha_{x0} = \alpha_{y0} = 0.
\end{equation}
It is arbitrarily chosen;
\begin{equation}
    \beta_x = \text{Max}, \hspace{2mm} \beta_y = \text{Min}.
\end{equation}
The focusing scheme is a periodic array of the following elements, constructing the FODO cell;
\begin{equation}
    [\text{Half-FQ}][\text{Drift}][\text{DQ}][\text{Drift}][\text{Half-FQ}].
\end{equation}
It is convenient to split the FODO cell into half FODO cells, where the split is at some point within the DQ such that $\alpha_{x1} = \alpha_{y1} = 0$. For zero acceleration, the split is at the mid-point. The drift sections are replaced with cavity sections, sandwiched between short drift lengths, of length $l_g$, to closer approximate a realistic beam line. The half focusing quadrupoles are described with length $l_{q1, 2}$ (where the second index (2) refers to the quadrupole being the second half of a complete quadrupole) and strength $k_1$, as shown in Fig.~\ref{fig:fodo_schem}. For now, the second quadrupole index is dropped ($l_{q1,2} = l_{q1}$).

The half FODO cell in the $x$ plane is thus;
\begin{equation}
    M = [\text{Half-FQ}][\text{Drift}][\text{Cavity}][\text{Drift}][\text{Half-DQ}].
\end{equation}

Explicitly, the transfer map is as follows;
\begin{multline}
\\
M = \begin{pmatrix}
    \cosh(\sqrt{k_2}l_{q2}) & \frac{1}{\sqrt{k_2}}\sinh(\sqrt{k_2}l_{q2})\\
    \sqrt{k_2}\sinh(\sqrt{k_2}l_{q2}) & \cosh(\sqrt{k_2}l_{q2})\\
    \end{pmatrix}\\
    \begin{pmatrix}
    1&  L_{eff} \\
    0 & \frac{\gamma_{r0}\beta_{r0}}{\gamma_{r1}\beta_{r1}}
    \end{pmatrix}\\
    \begin{pmatrix}
    \cos(\sqrt{k_1}l_{q1}) & \frac{1}{\sqrt{k_1}}\sin(\sqrt{k_1}l_{q1})\\
     -\sqrt{k_1}\sin(\sqrt{k_1}l_{q1}) & \cos(\sqrt{k_1}l_{q1})\\
    \end{pmatrix}.
    \label{eq:half_fodo_thick_map}
\end{multline}
Where the [\text{Drift}][\text{Cavity}][\text{Drift}] matrix have been multiplied together, and,
\begin{equation}
  \begin{aligned}
    L_{eff} = l_{g}\left(\frac{\gamma_{r0}\beta_{r0}}{\gamma_{r1}\beta_{r1}}+1\right)+\\ l_{cav}\frac{\gamma_{r0}\beta_{r0}}{\gamma_{r1} - \gamma_{r0}}ln\left(\frac{\gamma_{r1}\beta_{r1} + \gamma_{r1}}{\gamma_{r0}\beta_{r0} + \gamma_{r0}}\right).
  \end{aligned}
  \label{eq:Leff}
\end{equation}

Using Eqn.~\ref{eq:twiss_matrix} to transform the Twiss parameters due to transfer map M from $\alpha_{x0}=0$ to $ \alpha_{x1} = 0$,
\begin{equation}
    0 = -M_{11}M_{21}\beta_{x0} -\frac{M_{12}M_{22}}{\beta_{x0}}.
\end{equation}
\begin{comment}
and for y;
\begin{equation}
    0 = -M_{33}M_{43}\beta_{y0} -\frac{M_{34}M_{44}}{\beta_{y0}}
\end{equation}
\end{comment}
This produces an analytical form for the Twiss $\beta$ functions at the start of the half FODO cell, as functions of the half FODO transfer map, M;
\begin{equation}
    \beta_{x0} = \sqrt{\frac{-M_{12}M_{22}}{M_{11}M_{21}}}, \hspace{0.5cm}
    \beta_{y0} = \sqrt{\frac{-M_{34}M_{44}}{M_{33}M_{43}}}.
    \label{eq:initial_betas}
\end{equation}

Enforcing the beam size in $x$ at the start of the half FODO is equal to the $y$ beam size at the end of the half FODO;
\begin{equation}
    \sigma_{x0} = \sigma_{y1}, \hspace{0.5cm} \sigma_{x1} = \sigma_{y0}.
    \label{eq:equal_beam_size}
\end{equation}
The beam size can be determined with the following;
\begin{equation}
    \sigma = \sqrt{\frac{\beta\varepsilon_n}{\gamma_r\beta_r}},
\end{equation}
therefore Eqn.~\ref{eq:equal_beam_size} becomes
\begin{equation}
    \frac{\gamma_{r1}\beta_{r1}}{\gamma_{r0}\beta_{r0}}\beta_{x0} =  \beta_{y1}, \hspace{0.5cm} \frac{\gamma_{r1}\beta_{r1}}{\gamma_{r0}\beta_{r0}}\beta_{y0} =  \beta_{x1}.
    \label{eq:beta_x0/f=beta_y1}
\end{equation}
Therefore
\begin{equation}
    \beta_{x0}\beta_{x1} = \beta_{y0}\beta_{y1}.
    \label{eq:beta_relationship}
\end{equation}

The ratio of the $\beta$ functions at each half FODO cell, r, is given
\begin{equation}
    \frac{\beta_{x0}}{\beta_{y0}} = \frac{\beta_{y1}}{\beta_{x1}} = r.
\end{equation}
The Twiss $\beta$ and $\gamma$ functions at the end of the half FODO can be determined using Eqn.~\ref{eq:twiss_matrix};
\begin{equation}
    \beta_{x1} = {\frac{\gamma_{r1}\beta_{r1}}{\gamma_{r0}\beta_{r0}}}\left(M_{11}^2\beta_{x0} + \frac{M_{12}^2}{\beta_{x0}}\right)
\end{equation}
\begin{equation}
    \beta_{y1} = {\frac{\gamma_{r1}\beta_{r1}}{\gamma_{r0}\beta_{r0}}}\left(M_{33}^2\beta_{y0} + \frac{M_{34}^2}{\beta_{y0}}\right)
\end{equation}
\begin{equation}
    \gamma_{x1} = \frac{1}{\beta_{x1}} = {\frac{\gamma_{r1}\beta_{r1}}{\gamma_{r0}\beta_{r0}}}\left(M_{21}^2\beta_{x0} + \frac{M_{22}^2}{\beta_{x0}}\right)
\end{equation}
\begin{equation}
    \gamma_{y1} = \frac{1}{\beta_{y1}} = {\frac{\gamma_{r1}\beta_{r1}}{\gamma_{r0}\beta_{r0}}}\left(M_{43}^2\beta_{y0} + \frac{M_{44}^2}{\beta_{y0}}\right).
\end{equation}

Combining the above equations, along with Eqns.~\ref{eq:initial_betas}, \ref{eq:beta_relationship} and the equality $\det(M_x)$ = $\det(M_y)$, it can be shown;
\begin{equation*}
    M_{12} = \pm M_{34}, \hspace{0.5cm} M_{21} = \pm M_{43} \\
\end{equation*}
and
\begin{equation}
    M_{11}M_{22} = M_{33}M_{44}.
    \label{eq:relationships_M}
\end{equation}

Now that the basic relationships between half FODO cell elements have been determined, it is required to expand the elements as functions of quadrupole, drift length and cavity parameters. In order to proceed, quadrupole maps are simplified using the semi-thin lens approximation.

The semi-thin lens approximation expands trigonometric and hyperbolic functions and truncates all terms of the order $k^nl_{q1}^{n+2}$;
\begin{gather*}
    \cos(\sqrt{k_1}l_{q1}) \approx 1 - k_1l_{q1}^2/2\\
    \sin(\sqrt{k_1}l_{q1}) \approx \sqrt{k_1}l_{q1}\\
    \cosh(\sqrt{k_1}l_{q1}) \approx 1 + k_1l_{q1}^2/2\\
    \sinh(\sqrt{k_1}l_{q1}) \approx \sqrt{k_1}l_{q1}.
\end{gather*}
It also assumed the drift length between quadrupoles and cavities, $l_g << 1$ and therefore any terms of the order $k_1^nl_g^{n+2}$ are also ignored.

Substituting the semi-thin lens approximations into Eqn.~\ref{eq:half_fodo_thick_map} it is possible to show that the results in Eqn.~\ref{eq:relationships_M} can be simplified to the following identities;
\begin{equation}
    k_1l_{q1}^2 = k_2l_{q2}^2
    \label{eq:k1lq1^2=k2l_q2^2}
\end{equation}
\begin{equation}
    l_{q1} = \frac{\gamma_{r0}\beta_{r0}}{\gamma_{r1}\beta_{r1}}  l_{q2}
    \label{eq:lq1=fl_q2}
\end{equation}
\begin{equation}
    k_1  = \frac{\gamma_{r1}^2\beta_{r1}^2}{\gamma_{r0}^2\beta_{r0}^2} {k_2}
    \label{eq:k_1 = k_2/f^2}
\end{equation}
\begin{equation}
       \frac{\gamma_{r0}\beta_{r0}}{\gamma_{r1}\beta_{r1}}  k_1l_{q1}= k_2l_{q2}.
       \label{eq: f k_1l_q1 = k_2l_q2}
\end{equation}
For zero acceleration, Eqns.~\ref{eq:lq1=fl_q2}, \ref{eq:k_1 = k_2/f^2} and \ref{eq: f k_1l_q1 = k_2l_q2} return to the expected case. The above results are also solutions for full order quadrupole elements.

In order to find the minimum aperture possible for a given cavity length, the $\beta_x/\beta_y$ function at the start/end of the cavity is at a minimum. The transfer map, $\Lambda_x$, that transforms phase space from the initial position to the cavity entrance in $x$ is a (semi-thin lens) focusing quadrupole of length $l_{q1}$ followed by a drift of length $l_g$;
\begin{equation}
\\
\Lambda = 
    \begin{pmatrix}
    1 &  l_g \\
    0 & 1
    \end{pmatrix}\\
    \begin{pmatrix}
    1- k_1l_{q1}^2/2 & l_{q1}\\
     -k_1 &  1- k_1l_{q1}^2/2\\
    \end{pmatrix}.
    \label{eq:quad_drift_map}
\end{equation}

The $\beta$ function at the cavity entrance, $\beta_{xc0}$, is determined using Eqn.~\ref{eq:twiss_matrix} and~\ref{eq:quad_drift_map};
\begin{equation}
    \beta_{xc0} = \Lambda_{11}^2 \beta_{x0} + \frac{\Lambda_{12}^2}{\beta_{x0}}.
    \label{eq:beta_xc0}
\end{equation}
$\beta_{xc0}$ is minimised by differentiating Eqn.~\ref{eq:beta_xc0} with respect to quadrupole parameters, $k_1$, and equating to 0. It was found that differentiating with respect to $l_{q1}$ was not optimal, as produces quadrupole lengths of the order $~$1~m.

\begin{multline}
    \frac{d\beta_{xc0}}{dk_1} = 2\Lambda_{11}\frac{d\Lambda_{11}}{dk_1}\beta_{x0} + \Lambda_{11}^2 \frac{d\beta_{x0}}{dk_1} + \\  2\Lambda_{12}\frac{d\Lambda_{12}}{dk_1}\frac{1}{\beta_{x0}} - \Lambda_{12}^2 \frac{1}{\beta_{x0}^2} \frac{d\beta_{0}}{dk_1} = 0.
\end{multline}
Where the total derivative is taken. Rearranging for the derivative of $\beta_{x0}$ with respect to $k_1$;
\begin{equation}
    \frac{d\beta_{x0}}{dk_1} = \frac{-2\Lambda_{11}\beta_{x0}\frac{d\Lambda_{11}}{dk_1} - 2\Lambda_{12}\beta_{x0}^{-1}\frac{d\Lambda_{12}}{dk_1}}{\Lambda_{11}^2 - \Lambda_{12}^2\beta_{x0}^{-2}}.
    \label{eq:dbeta/dk_1}
\end{equation}
A form for $\beta_{x0}$ can be computed in the semi-thin lens regime using Eqn.~\ref{eq:initial_betas}. The result is;
\begin{equation}
    \beta_{x0} \approx \frac{\sqrt{r}}{k_1l_{q1}}\sqrt{1 + \frac{l_{q1}}{L_{eff,1}}},
    \label{eq:beta_x0_semi_thin}
\end{equation}
which is subsequently differentiated with respect to $k_1$,
\begin{equation}
    \frac{d\beta_{x0}}{dk_1} =\beta_{x0}\left(\frac{1}{2r}\frac{dr}{dk_1} - \frac{1}{k_1}\right).
    \label{eq:dbeta/dk_2}
\end{equation}

The aspect ratio, $r$, can be expanded in the semi-thin lens regime,
\begin{equation}
    r = \frac{M_{33}}{M_{11}} \approx \frac{1 +L_{eff,1}k_1l_{q1} + k_1l_{q1}^2 -\frac{L_{eff,1}k_1^2l_{q1}^3}{2}}{1 - L_{eff,1}k_1l_{q1} - k_1l_{q1}^2 - \frac{L_{eff,1}k_1^2l_{q1}^3}{2}},
    \label{eq:r_semi_thin}
\end{equation}
which can be differentiated with respect to $k_1$, as required in Eqn.~\ref{eq:dbeta/dk_2}. Combining Eqns.~\ref{eq:quad_drift_map}, ~\ref{eq:dbeta/dk_1},~\ref{eq:dbeta/dk_2},~\ref{eq:beta_x0_semi_thin}, and Eqn.~\ref{eq:r_semi_thin} before simplifying and ignoring all terms smaller than the semi-thin lens limit, produces a cubic in $k_1$;

\begin{multline}
  \label{eq:cubic_in_k}
    -\frac{l_{q1}^4L_{eff}^2}{2} k_1^3 + l_{q1}^2(2l_g L_{eff} -2l_{q1}L_{eff} - L_{eff}^2)k_1^2  \\ -l_{q1} (L_{eff} +l_{q1} )k_1 +1  = 0.
\end{multline}
It can be shown that for reasonable values for $l_{q1}, k_1, L_{eff}$, Eqn.~\ref{eq:cubic_in_k} has three real roots, and thus trigonometric solutions exist \cite{cubic}. The solutions are as follows, for $m = 0, 1, 2$.

\begin{equation}
    k_1 = 2\sqrt{\frac{-p}{3}} \cos\left(\arccos\left(\frac{3q}{2p}\right)\sqrt{\frac{-1}{3p}} - \frac{2\pi m}{3}\right) -\frac{b}{3a},
    \label{eq:k1_semi_thin}
\end{equation}
where
\begin{equation*}
    p = \frac{3ac -b^2}{3a^2} ,\hspace{0.1cm}
    q = \frac{2b^3 - 9abc + 27a^2}{27a^3}
\end{equation*}
and 
\begin{gather*}
    a = \frac{-L_{eff,1}^2 l_{q1}^4}{2} \\
    b = l_{q1}^2L_{eff,1}(2(l_g -l_{q1}) -L_{eff,1})\\
    c = -l_{q1}(L_{eff,1} + l_{q1}).
\end{gather*}
Equation~\ref{eq:cubic_in_k} can also be solved using the thin lens approximation, keeping terms of the form $k_1^nl_{q1}^n$. The result is;
\begin{equation}
    k_1 = \frac{1}{l_{q1}L_{eff,1}}\frac{\sqrt{5}-1}{2}.
    \label{eq:k1_thin}
\end{equation}
Equations~\ref{eq:k1_semi_thin} and \ref{eq:k1_thin} produce analytical methods to determine the optimum value of $k_1$ such that the maximum transverse beam size is minimised at the cavity entrance/exit, for a given cavity length and quadrupole length, within a FODO-like scheme.

\begin{figure}[h]
\centering
    \subcaptionbox{}{\includegraphics[width=3in]{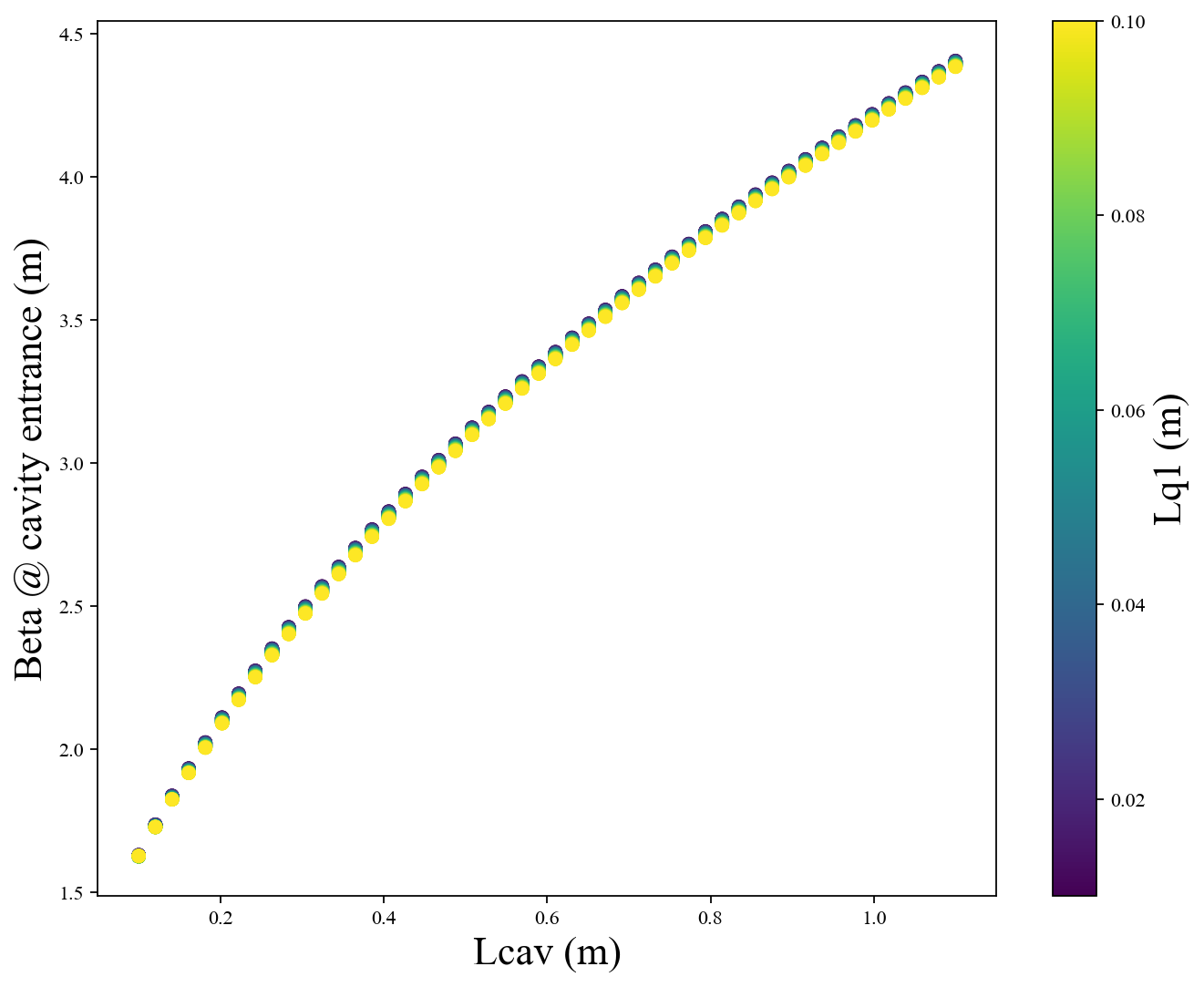}}
     \caption{Optimum value of $\beta_{xc0}$ as a function of cavity length and first quadrupole length.}
     \label{fig:beta_xc0_func_lq1}
\end{figure}

It can be shown the value of $\beta_{xc0}$ is relatively insensitive to $l_{q1}$. As a result, the user defined value of $l_{q1}$ is not heavily constrained. However, as the semi-thin lens regime is adopted, $l_{q1}$ can not approach similar values to $l_{cav}$. Figure~\ref{fig:beta_xc0_func_lq1} displays $\beta_{xc0}$ as a function of cavity length and $l_{q1}$. The value of $k_1$ is calculated with Eqn.~\ref{eq:k1_semi_thin}. For longer cavity lengths, $\beta_{xc0}$ is larger, as expected. The value of $\beta_{xc0}$ is highly insensitive to initial values of $l_{q1}$.

\subsection{Semi-thin lens relative to thin lens}

\begin{figure}[h]
\centering
    \subcaptionbox{}{\includegraphics[width=3in]{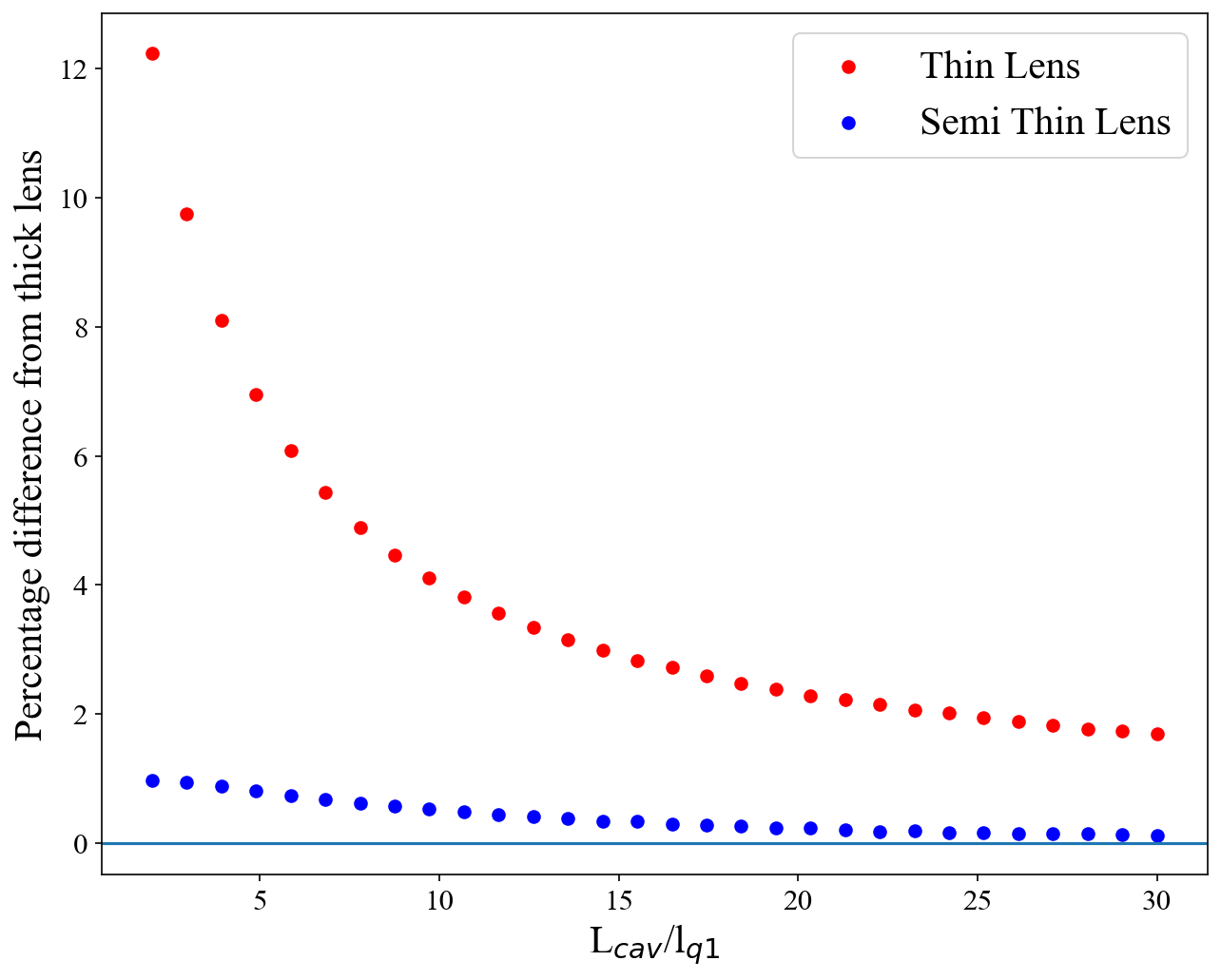}}
     \caption{The percentage difference between optimal value of $k_1$ as calculated by thick lens and the thin and semi-thin regimes. $l_{q1}$ = 0.05 m, $l_{g}$= 0.05 m.}
     \label{fig:semi_thin_error}
\end{figure}

Figure~\ref{fig:semi_thin_error} displays the percentage difference between the optimal value of $k_1$ as calculated by the thick lens regime (solved numerically) and the thin/semi-thin regimes. For the minimum ratio $L_{cav}/l_{q1}$ = 2, the percentage difference between the semi-thin and thick lens regime is less than 1$~\%$, approximately 10$\times$ less than the thin lens regime. Whilst the accuracy of the semi-thin lens regime is a function of $L_{cav}/l_{q1}$, the error increases with $l_{q1}$, for the same value of $L_{cav}/l_{q1}$.

\section{Concatenating multiple half-FODO cells}
As there is nothing special about the first half FODO cell used to derive important constraints, the constraints extend to all half-FODO cells in a lattice, allowing for propagating equations to be formed. Firstly the second index describing the quadrupole length are reintroduced, describing if the quadrupole is the first or second half of the complete quadrupole unit, recall;
\begin{equation*}
    l_{q1} \rightarrow l_{q1,2},  \hspace{0.2cm} l_{q2} \rightarrow l_{q2,1}.
\end{equation*}
For a set of $N$ half FODO cells, there exists 2$N$ half quadrupoles. The $k$-strengths behave as follows
\begin{multline}
    k_1 = \frac{k_2}{\left( \frac{\gamma_{r0}\beta_{r0}}{\gamma_{r1}\beta_{r1}}\right)^2} = \frac{k_3}{\left( \frac{\gamma_{r0}\beta_{r0}}{\gamma_{r1}\beta_{r1}}\right)^2} =\\ \frac{k_4}{\left( \frac{\gamma_{r0}\beta_{r0}}{\gamma_{r1}\beta_{r1}}\right)^2 \left( \frac{\gamma_{r1}\beta_{r1}}{\gamma_{r2}\beta_{r2}}\right)^2} =\cdots= \frac{k_{2N}}{\prod_{i=0}^{N-1} \left( \frac{\gamma_{ri}\beta_{ri}}{\gamma_{r(i+1)}\beta_{r(i+1)}}\right)^2}.
    \label{eq:k_concat}
\end{multline}
Where it was used that $k_{2N}$ = $k_{2N+1}$ as they are two sections of the same quadrupole but separated into two half FODO cells.
From Eqn.~\ref{eq: f k_1l_q1 = k_2l_q2}, the relationship between consecutive quadrupole lengths (first section) can also be determined;
\begin{equation}
  l_{q(n),1} = \frac{\gamma_{r(n-1)}\beta_{r(n-1)}}{\gamma_{r0}\beta_{r0}} l_{q1,2}.
  \label{eq:lq_concat}
\end{equation}
By defining $l_{q1,2}$ and values for the Lorentz factor, all quadrupole $k$-strengths and first section lengths can be determined. The second section lengths of quadrupoles must now be determined.

In order to satisfy the constraint in Eqn.~\ref{eq:beta_x0/f=beta_y1}, Eqn.~\ref{eq:beta_x0_semi_thin} is combined with the fact
\begin{equation}
    \beta_{y1} \approx \frac{\sqrt{r}}{k_2l_{q2,2}}\sqrt{1 + \frac{l_{q2,2}}{L_{eff,2}}},
\end{equation}
producing a constraint on values for $L_{eff, n}$;
\begin{equation}
    L_{eff,2} = \frac{\gamma_{r1}\beta_{r1}}{\gamma_{r0}\beta_{r0}}{L_{eff,1}}.
    \label{eq:Leff_constraint}
\end{equation}
As a constant aspect ratio was assumed, this constraint must be enforced
\begin{multline}
    r_1 \approx \frac{1 +L_{eff,1}k_1l_{q1,2} + k_1l_{q1,2}^2 -\frac{L_{eff,1}k_1^2l_{q1,2}^3}{2}}{1 - L_{eff,1}k_1l_{q1,2} - k_1l_{q1,2}^2 - \frac{L_{eff,1}k_1^2l_{q1,2}^3}{2}} \approx \\ r_2 \approx \frac{1 +L_{eff,2}k_2l_{q2,2} + k_2l_{q2,2}^2 -\frac{L_{eff,2}k_2^2l_{q2,2}^3}{2}}{1 - L_{eff,2}k_2l_{q2,2} - k_2l_{q2,2}^2 - \frac{L_{eff,2}k_2^2l_{q2,2}^3}{2}}.
    \label{eq:constant_r}
\end{multline}
Substituting with Eqns.~\ref{eq:k_1 = k_2/f^2} and~\ref{eq:Leff_constraint}, it is a requirement that
\begin{equation}
    l_{q2,2} ={\frac{\gamma_{r1}\beta_{r1}}{\gamma_{r0}\beta_{r0}}}  {l_{q1,2}}.
    \label{eq:lq_concat_2}
\end{equation}
From Eqn.~\ref{eq:lq_concat}, $l_{q2,1} = l_{q2,2}$. This result; quadrupole sections of the same quadrupole unit are the same length (in addition to $k$-strength), and the maximum/minimum beam size occurs at the centre point of the quadrupole unit.

In order to satisfy Eqn.~\ref{eq:Leff_constraint}, either the drift or cavity length (or a combination of the both) can be altered within consecutive half-FODO cells (see Eqn.~\ref{eq:Leff}). The required change in element length manifests differently in each elements. As drift lengths are short relative to cavity lengths, the drifts become long, and the real estate gradient drops. When the constraint term is absorbed by increasing consecutive cavity lengths, the additional length does not cause a drop in real estate gradient. In fact, it can be shown that a FODO-like scheme is possible such that cavity lengths increase faster than quadrupole lengths, thus producing a lattice with higher real estate gradient than the standard FODO scheme. Figure~\ref{fig:solving_Leff_for_different_lattice} shows the change in real estate gradient for different methods to solve Eqn.~\ref{eq:Leff_constraint}.

In this section, the FODO-like focusing scheme was explored. Given an initial set of parameters, namely the first quadrupole length, drift length and cavity length, the value of all quadrupole lengths and $k$-strengths are determinable, such that the limiting beam size is minimised at the cavity entrance/exit. In addition to the quadrupole parameters, consecutive cavity lengths and drift lengths are constrained such that the aspect ratio and beam size are constant at each half FODO cell.

\section{FODO Results}

\begin{figure}[h]
\centering
     \subcaptionbox{}{\includegraphics[width=3in]{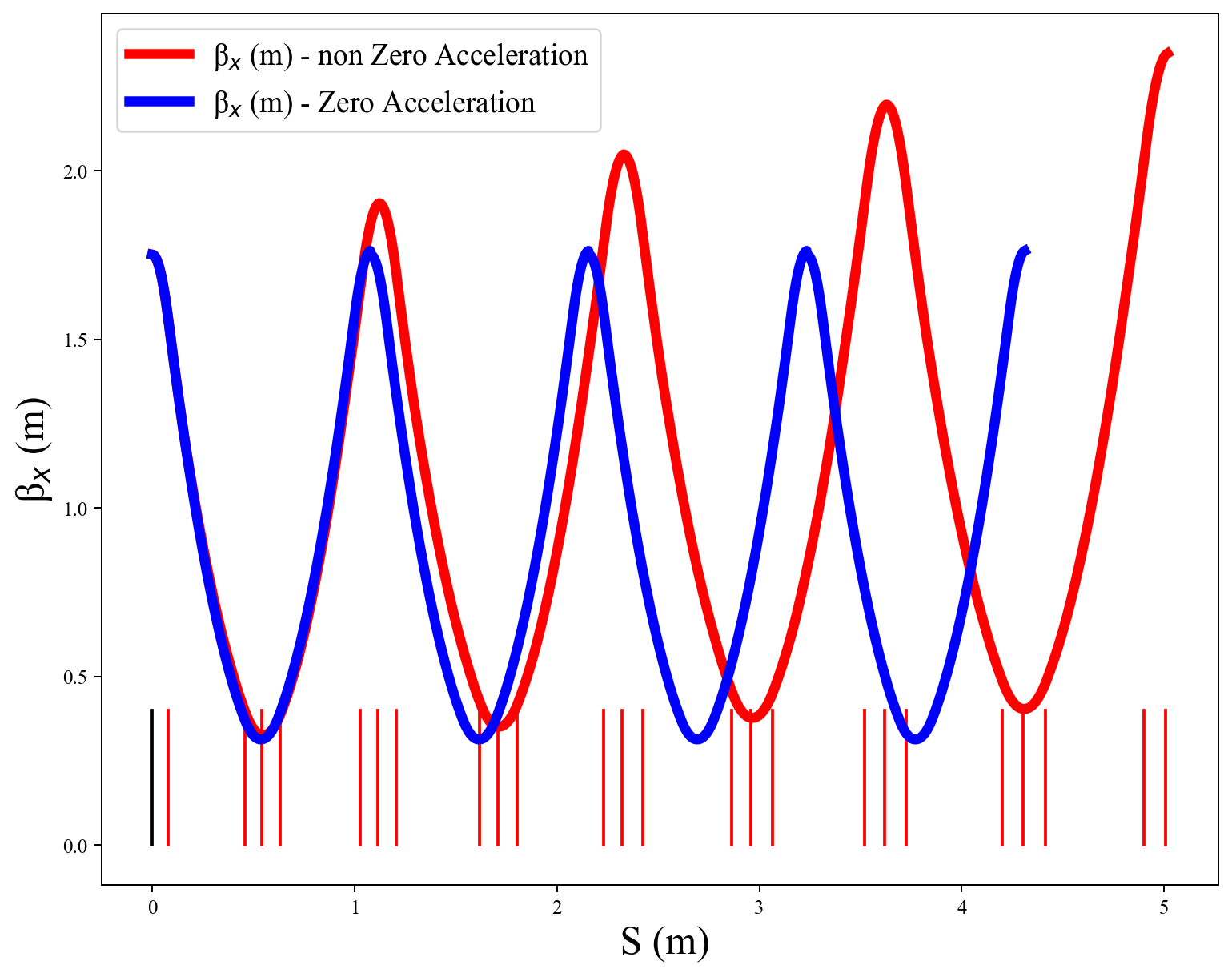}}
     \subcaptionbox{}{\includegraphics[width=3in]{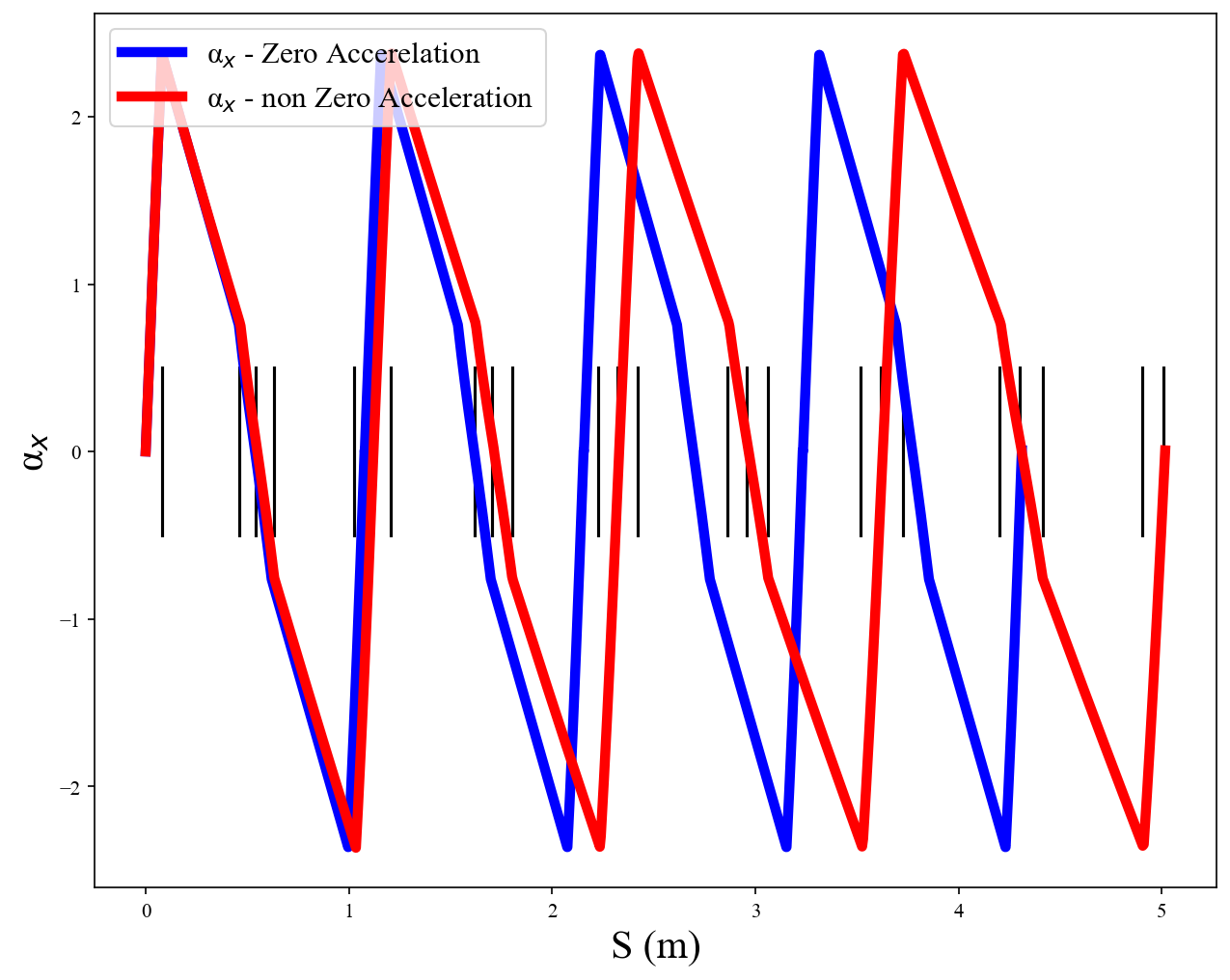}}
     \caption{Twiss $\beta$ (a) and $\alpha$ (b) functions along FODO-like lattice.}
     \label{fig:Twiss_params}
\end{figure}

Figure~\ref{fig:Twiss_params} displays the $\beta_x$ (a) and $\alpha_x$ (b) Twiss parameters as a function of longitudinal displacement, $s$, over four FODO cells. Twiss parameters are shown for both the standard (zero acceleration) and FODO-like (non-zero acceleration) lattice. The standard FODO refers to the case of constant quadrupole strengths and lengths along the lattice, in addition to the drift lengths. The FODO-like lattice refers to the case where lattice parameters change as describe by Eqns.~\ref{eq:k_concat}, \ref{eq:lq_concat}, \ref{eq:Leff_constraint} and \ref{eq:lq_concat_2}. The lattice is comprised of cavities with lengths of the order 1~m and gradients of 50 MeV/m. $l_{q1}$ = 0.01~m and $l_{g1} =~$0.05~m. The maximum $\beta_x$ function for a FODO-like lattice increases with $s$, as the Lorentz factor increases due to acceleration from rf cavities.

\begin{figure}[h]
\centering
     \subcaptionbox{}{\includegraphics[width=3in]{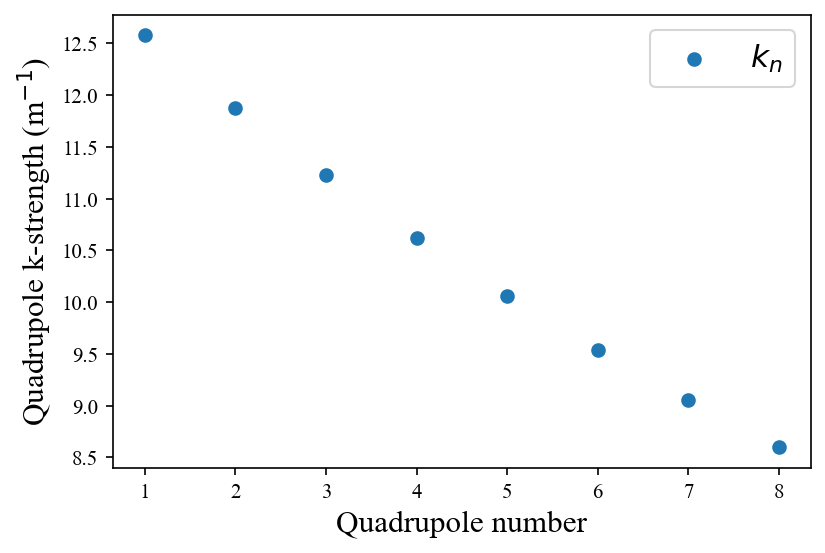}}
     \subcaptionbox{}{\includegraphics[width=3in]{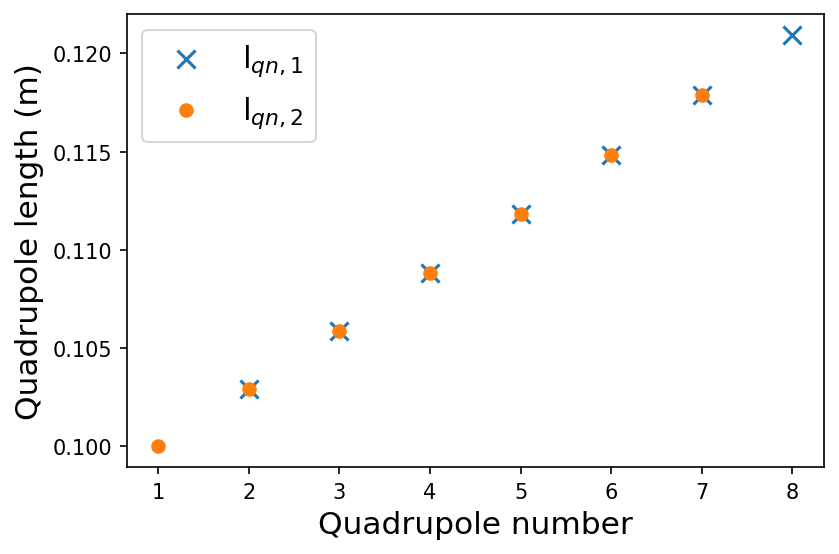}}
     \caption{Quadrupole $k$-strength (a) and length (b) as a function of quadrupole number in FODO-like scheme.}
     \label{fig:Quad_params}
\end{figure}

\begin{figure}[h]
\centering
    \subcaptionbox{}{\includegraphics[width=3in]{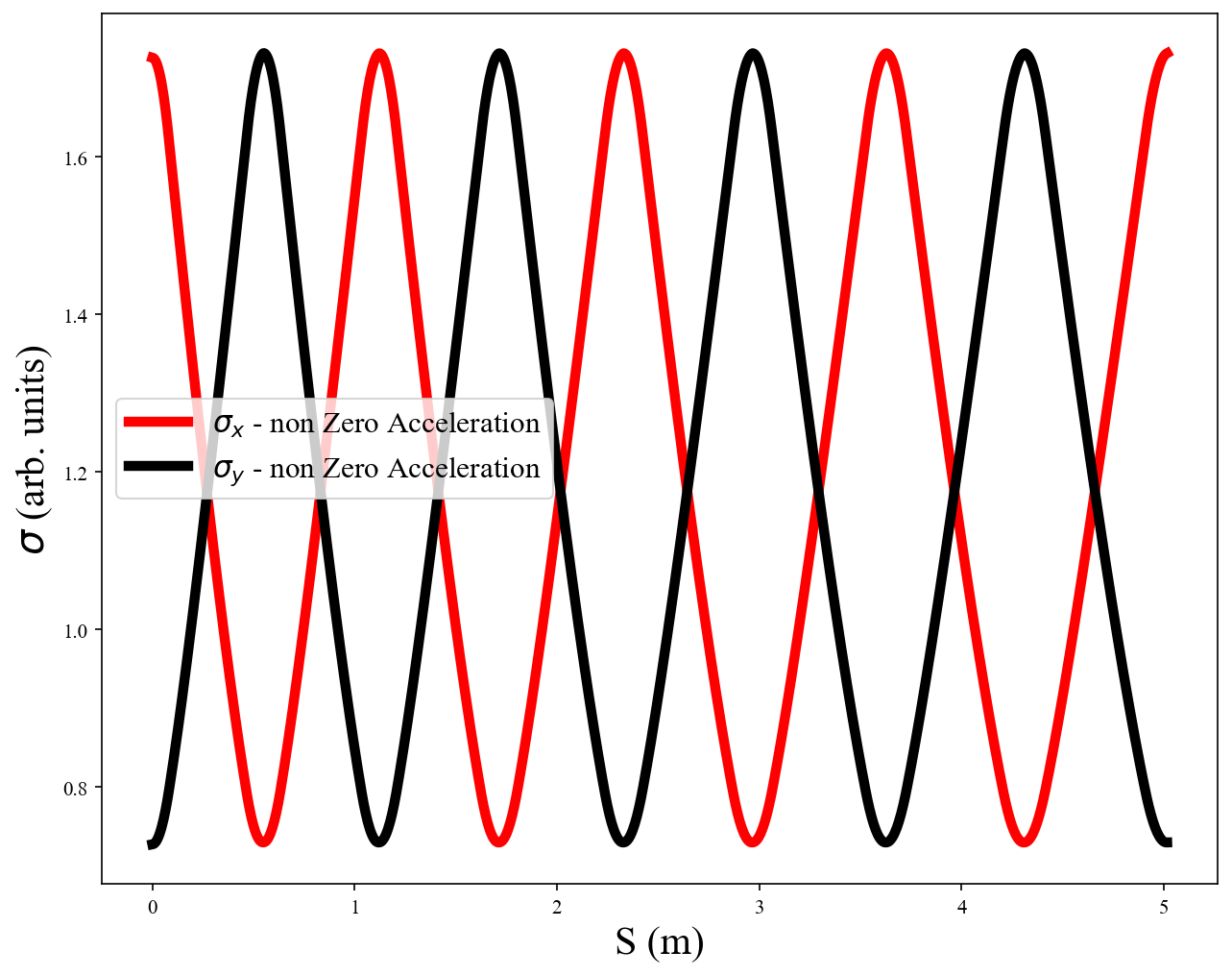}}
     \caption{Transverse beam size in the FODO-like scheme.}
     \label{fig:beam_size}
\end{figure}

The total length of the FODO lattice is longer for the FODO-like lattice, as the quadrupole lengths, cavity lengths and/or drift lengths increase with Lorentz factor, from Eqns.~\ref{eq:lq1=fl_q2} and~\ref{eq:Leff_constraint}. Figure~\ref{fig:Quad_params} demonstrates the decrease/increase in consecutive quadrupole $k$-strength/lengths along a FODO-like lattice. The constant beam size in both transverse planes along a FODO-like lattice are shown in Fig.~\ref{fig:beam_size}, as required.

\begin{figure}[h]
\centering
    \subcaptionbox{}{\includegraphics[width=3in]{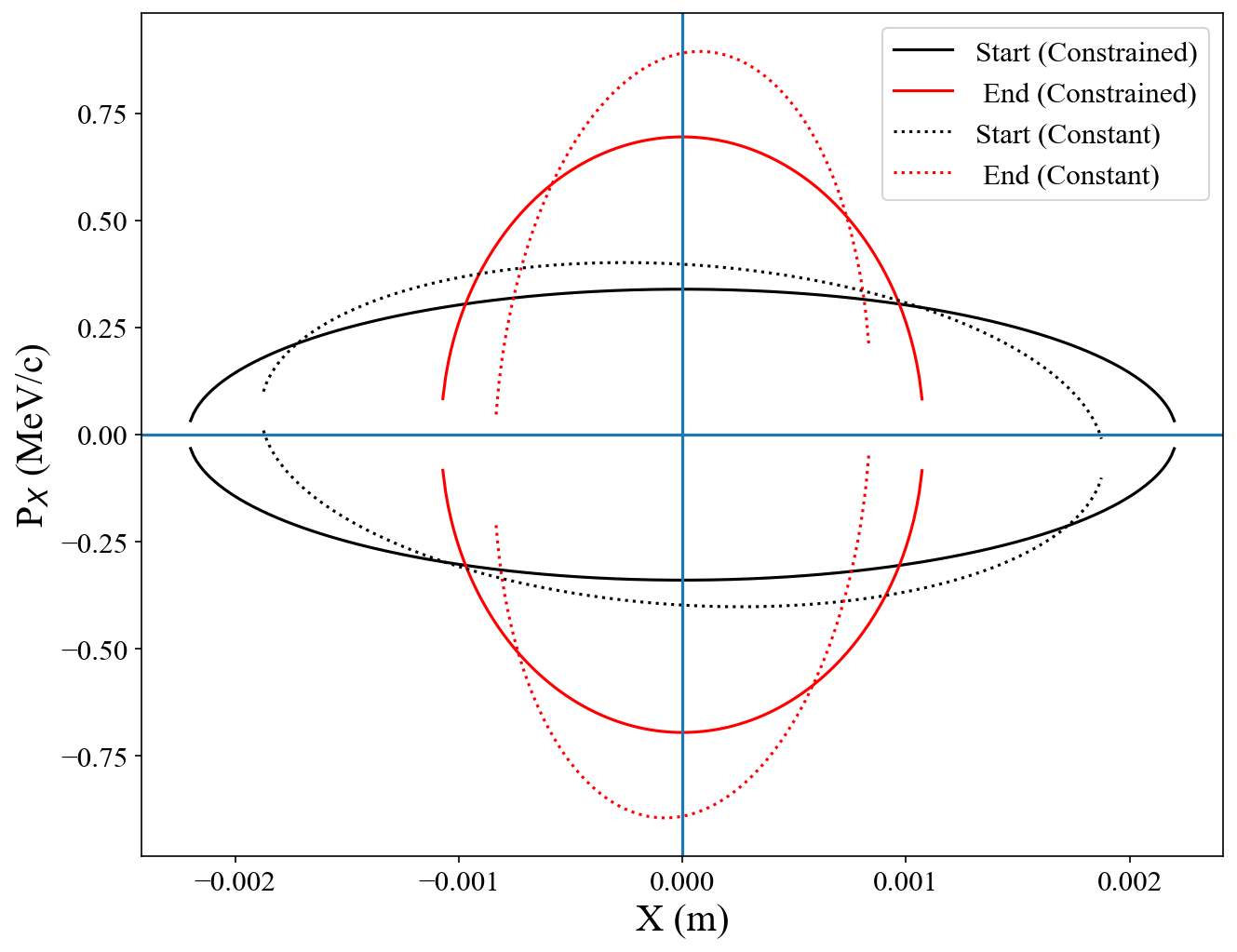}}
     \caption{Phase space ellipse at fifth half FODO cell calculated with constant or constrained lattice parameters.}
     \label{fig:beam_ellipse_fodo_lattice}
\end{figure}

Figure~\ref{fig:beam_ellipse_fodo_lattice} displays the $x$ phase space ellipse at the entrance of the fifth half FODO cell as calculated by both a constant (standard FODO) and constrained (FODO-like) FODO lattice. For constant lattice parameters, the phase space ellipse is over/under focused at half FODO cell boundaries, as the constant aspect ratio and beam size constraint are not met. The constrained lattice produces a well matched ellipse at the boundary, as required. In the limit of a high number of periodic FODO cells, the standard FODO scheme remains stable, when acceleration is incorporated.

The standard FODO lattice produces larger real estate gradients to the FODO-like lattice when considering long and fixed cavity lengths. In this case, the drift length must increase along the lattice to keep Eqn.~\ref{eq:Leff_constraint} satisfied. As the cavity lengths are long, the correction to subsequent $l_g$ is large, and the real estate gradient drops. In addition, increased $l_g$ causes the beam to defocus longitudinally.

\begin{figure}[h]
\centering
     \subcaptionbox{}{\includegraphics[width=3in]{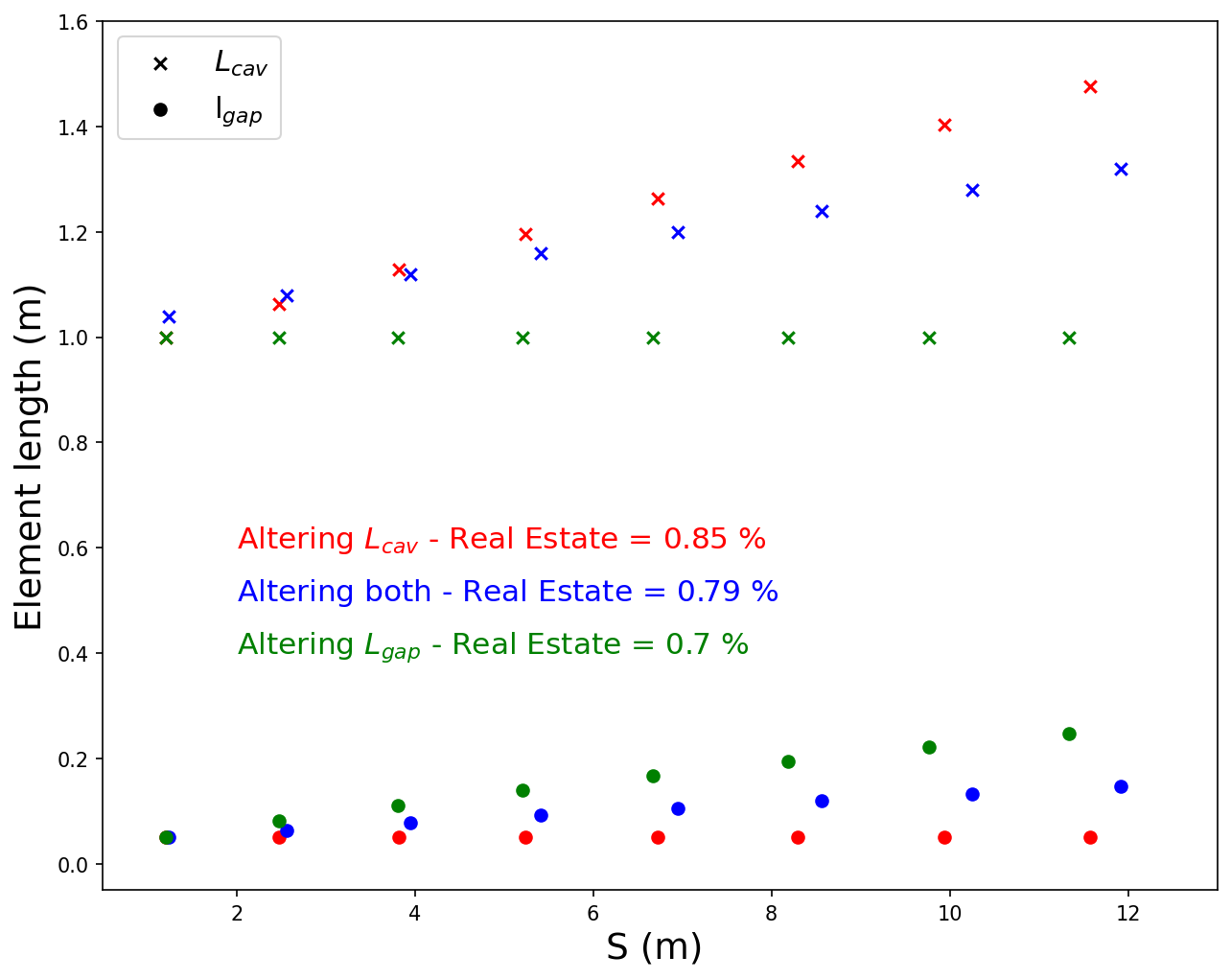}}
     \caption{The development of the cavity and drift length, for different methods of satisfying Eqn.~\ref{eq:Leff_constraint}.}
     \label{fig:solving_Leff_for_different_lattice}
\end{figure}

For linacs with short cavity lengths, the correction absorbed by $l_g$ is small, and the defocusing effect is suppressed. Thus for short cavity length, The FODO-like scheme becomes an effective focusing scheme. As previously discussed, cavity lengths can be defined to increase in length, such that the FODO-like lattice has higher real estate to the standard FODO lattice. Figure~\ref{fig:solving_Leff_for_different_lattice} displays the real estate gradient of the three different FODO-like lattice adopting different methods to satisfy Eqn.~\ref{eq:Leff_constraint}. It can be seen the maximum real estate gradient is achieved by keeping $l_g$ constant and increasing $L_{cav}$.

\section{\label{sec:MAS}Minimum Aperture Scheme}

\begin{figure}[h]
\centering
    \subcaptionbox{}{\includegraphics[width=3in]{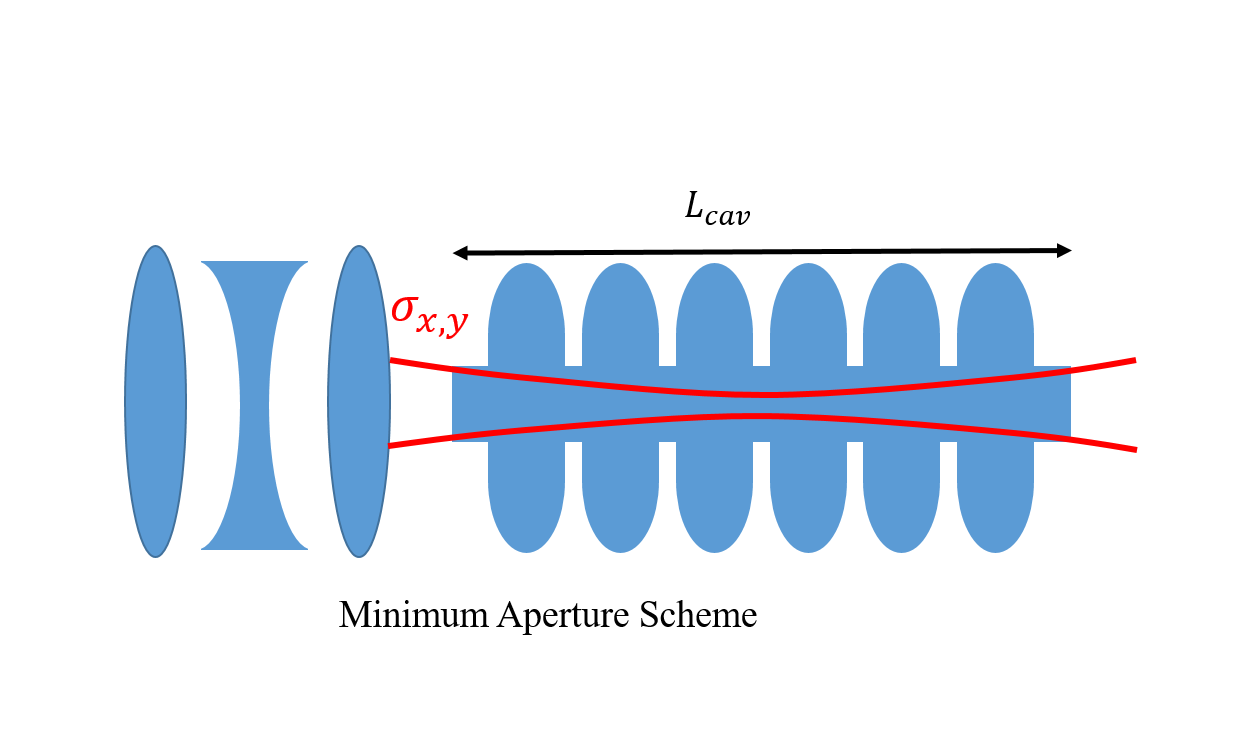}}
    \subcaptionbox{}{\includegraphics[width=3in]{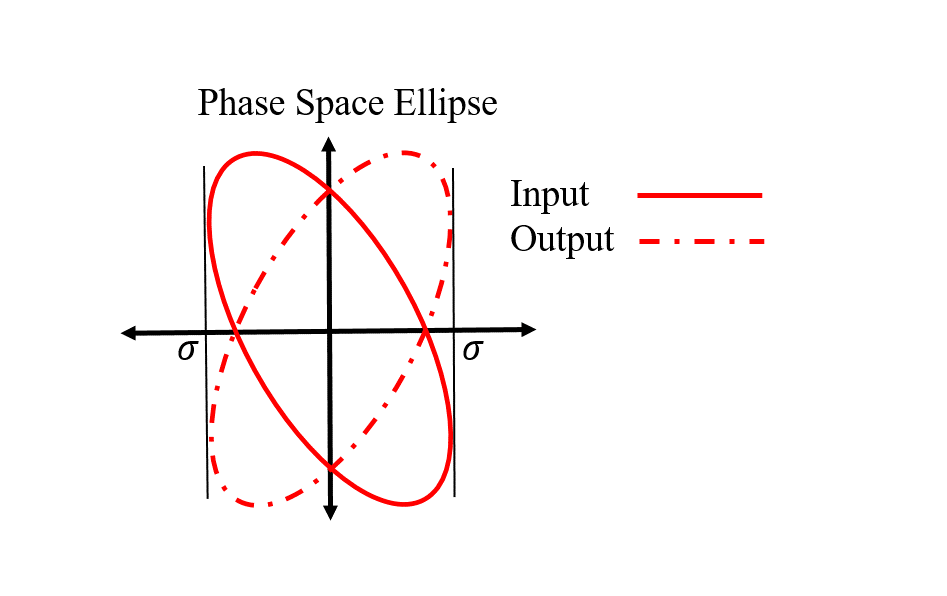}}
     \caption{(a) MAS Schematic. (b) Input/Output phase space ellipse in MAS.}
     \label{fig:mas_schematic}
\end{figure}

The MAS considers the focusing scheme by which a set of focusing elements are placed upstream of a cavity that orients the beam ellipse to match the acceptance ellipse of the cavity. For a given cavity length, the MAS produces the minimum cavity aperture that can be realised, for a given transverse beam emittance. A schematic of the MAS is shown Fig.~\ref{fig:mas_schematic}.

The MAS scheme realises the case of minimum beam size at the cavity entrance/exit. The first constraint thus forces equal beam size either side of the cavity;
\begin{equation}
    \sigma_{xc0} = \sigma_{xc1}, \hspace{2mm} \rightarrow \beta_{xc0} \frac{\gamma_{r1}\beta_{r1}}{\gamma_{r0}\beta_{r0}} = \beta_{xc1}. 
\end{equation}
The input beta function, $\beta_{xc0}$, is a constrained value given by the beam emittance and aperture size. $\beta_{xc1}$ is determinable using the Twiss parameter transform matrix (Eqn.~\ref{eq:twiss_matrix});
\begin{equation}
    \beta_{xc1} = \frac{\gamma_{r1}\beta_{r1}}{\gamma_{r0}\beta_{r0}}(R_{11}^2 \beta_{xc0} -2 R_{11}R_{12} \alpha_{xc0} + R_{12}^2 \gamma_{xc0}).
    \label{eq:beta_xc1_mas}
\end{equation}
The matrix elements, $R$ are defined by the rf cavity map shown in Eqn.~\ref{eq:cavity_map_simple}. Solving Eqn.~\ref{eq:beta_xc1_mas} for $\alpha_{xc0}$ produces a quadratic, solved using the quadratic formula. As there is only one set of Twiss parameters that can produce the required beam ellipse, the determinant must be zero. The results are shown below;
\begin{equation}
    \beta_{xc0} = L_{cav} \frac{\gamma_{r0}\beta_{r0}}{\gamma_{r1}-\gamma_{r0}} \ln\left(\frac{\gamma_{r1}\beta_{r1} + \gamma_{r1}}{\gamma_{r0}\beta_{r0} + \gamma_{r0}}\right) \approx L_{cav},
    \label{eq:mas_beta_xc0}
\end{equation}
\begin{equation}
    \alpha_{xc0} = 1,
    \label{eq:alpha_xc0}
\end{equation}
and,
\begin{equation}
    \gamma_{xc0} = \frac{2}{L_{cav} \frac{\gamma_{r0}\beta_{r0}}{\gamma_{r1}-\gamma_{r0}} \ln\left(\frac{\gamma_{r1}\beta_{r1} + \gamma_{r1}}{\gamma_{r0}\beta_{r0} + \gamma_{r0}}\right)}.
    \label{eq:gamma_xc0}
\end{equation}

It can be shown that for a given transverse emittance, the minimum cavity aperture as a function of cavity length is;
\begin{equation}
    a = \sqrt{\varepsilon_n\left(\frac{L_{cav}}{\gamma_{r1}-\gamma_{r0}} \ln\left(\frac{\gamma_{r1}\beta_{r1}+\gamma_{r1}}{\gamma_{r0}\beta_{r0}+\gamma_{r0}}\right)\right)}.
\end{equation}

\begin{figure}[h]
\centering
    \subcaptionbox{}{\includegraphics[width=3in]{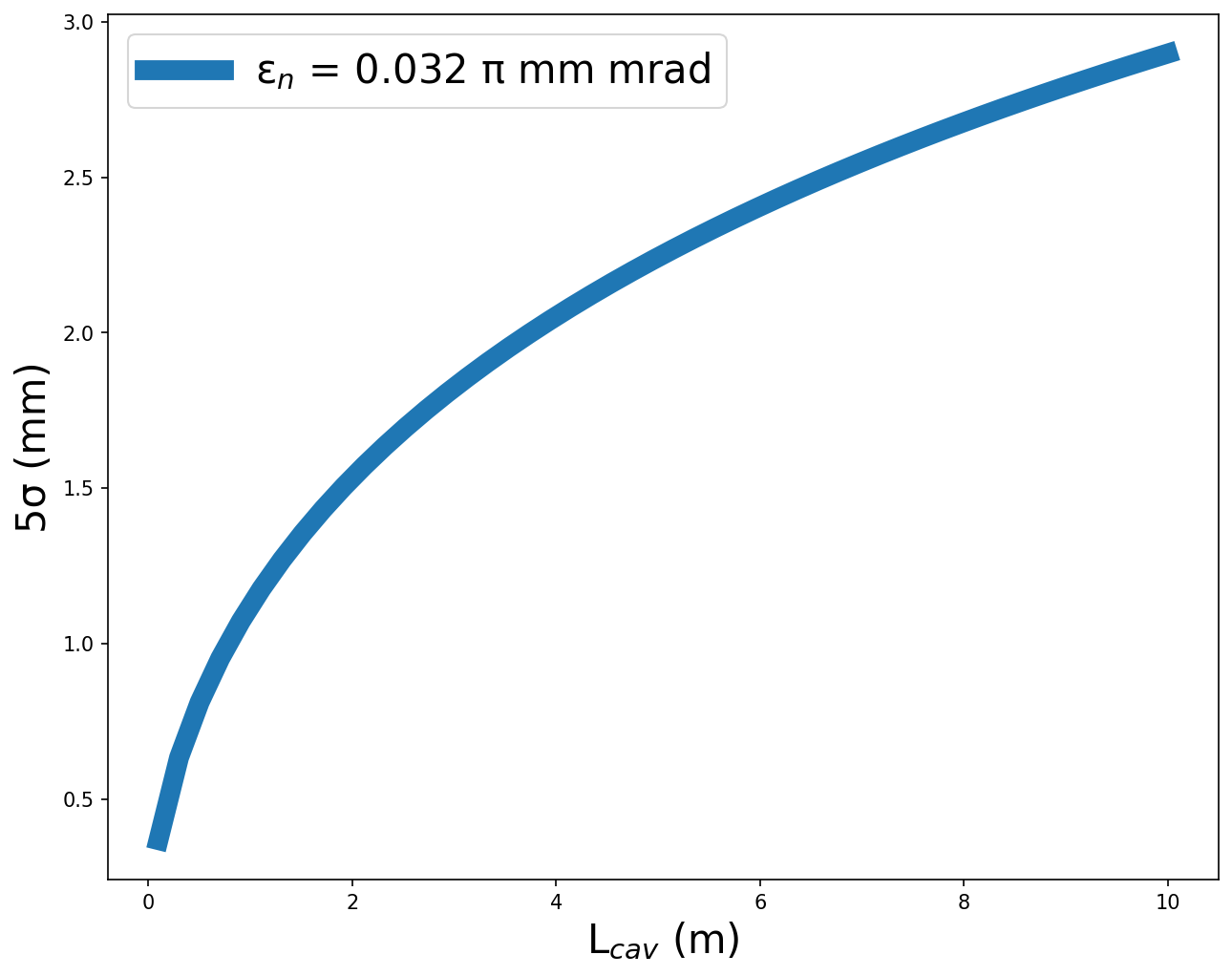}}
     \caption{Beam aperture as a function of cavity length in the MAS scheme.}
     \label{fig:aperture_limit_mas}
\end{figure}

Figure~\ref{fig:aperture_limit_mas} displays the minimum cavity aperture for a 5$\sigma$ beam as a function of cavity length with gradient of 50 MeV/m.

The MAS requires multiple quadrupoles to produce the required matching for the beam ellipse, and thus uses more quadrupole per cavity than the FODO-like scheme. However, the MAS can produce optimal focusing schemes, where the increase in non-active length (quadrupoles, drift lengths) is less than the increase in active cavity length. 

\section{Conclusion}
\label{section:conclusion}

In this paper, a self-consistent framework was demonstrated, that allowed the incorporation of acceleration into transverse beam dynamics studies for a proton linac machine. Two focusing schemes were developed and discussed; the FODO-like scheme, and the minimum aperture scheme. The FODO-like scheme is a simple scheme, requiring only one quadrupole per cavity. The scheme was analytically solved to minimise the beam size at the cavity entrance/exit and ensures constant beam size along the lattice. It was shown that lattice parameters must be altered along the FODO cell, to meet the design constraints for an accelerating scheme. The MAS describes the regime that matched the beam ellipse to the acceptance ellipse of a cavity, allowing for the smallest possible aperture, for a given cavity length. The MAS will require more than one quadrupole per cavity, and therefore will only have higher real estate gradients than the FODO-like scheme in special cases.

\begin{acknowledgments}
The studies presented have been funded through the Cockcroft Core Grant by STFC Grants No. ST/P002056/1 and ST/V001612/1.
\end{acknowledgments}

%\bibliography{bib}% Produces the bibliography via BibTeX.
%\printbibliography[title=References]
\bibliographystyle{plain} % We choose the "plain" reference style
\bibliography{main} % Entries are in the refs.bib file
\end{document}